\documentclass[superscriptaddress,groupedaddress]{revtex4}  % for review and submission
\usepackage{graphicx}  % needed for figures
\usepackage{dcolumn}   % needed for some tables
\usepackage{bm}        % for math
\usepackage{amssymb,amsmath,amsopn,amsfonts}
\usepackage{colordvi}
\usepackage{color}
\begin{document}

% The following information is for internal review, please remove them for submission
\widetext
%\leftline{Version xx as of \today}
%\leftline{Primary authors: Joe E. Physics}
%\leftline{To be submitted to PRL.}
%\leftline{Comment to {\tt d0-run2eb-nnn@fnal.gov} by xxx, yyy}
%\centerline{\em D\O\ INTERNAL DOCUMENT -- NOT FOR PUBLIC DISTRIBUTION}

% the following line is for submission, including submission to the arXiv!!
%\hspace{5.2in} \mbox{Fermilab-Pub-04/xxx-E}

\title{Robust manipulation of superconducting qubits in the presence of fluctuations}
%\input author_list.tex       % D0 authors (remove the first 3 lines
                             % of this file prior to submission, they
                             % contain a time stamp for the authorlist)
                             % (includes institutions and visitors)
\author{Daoyi Dong$^{1, \star}$,
Chunlin Chen$^{2, \star}$, Bo Qi$^{3}$, Ian R. Petersen$^{1}$, Franco Nori$^{4,5}$\\
$^{1}$\textit{School of Engineering and Information Technology, University of New South Wales, Canberra 2600, Australia}\\
$^{2}$\textit{Department of Control and System Engineering, Nanjing University, Nanjing 210093, China}\\
$^{3}$\textit{Key Laboratory of Systems and Control, ISS, and National Center for Mathematics and Interdisciplinary Sciences,
Academy of Mathematics and Systems Science, CAS, Beijing 100190, P. R. China}\\
$^{4}$\textit{CEMS, RIKEN, Wako-shi, Saitama, 351-0198 Japan}\\
$^{5}$\textit{Department of Physics, University of Michigan, Ann Arbor, Michigan 48109-1040 USA}\\
$^\star$\textit{email: daoyidong@gmail.com, clchen@nju.edu.cn}
}

\date{\today}

\begin{abstract}
Superconducting quantum systems are promising
candidates for quantum information processing due to their
scalability and design flexibility. However, the
existence of defects, fluctuations, and inaccuracies is unavoidable for
practical superconducting quantum circuits. In this paper, a
sampling-based learning control (SLC) method is used to guide the
design of control fields for manipulating superconducting quantum
systems. Numerical results for one-qubit systems and coupled
two-qubit systems show that the ``smart" fields learned using the
SLC method can achieve robust manipulation of superconducting
qubits, even in the presence of large fluctuations and
inaccuracies.
\end{abstract}

%\pacs{03.65.Wj, 02.50.-r, 03.67.-a}

\maketitle

%\section{\label{sec:level1}First-level heading}
% sections are not used for PRL papers

Superconducting quantum circuits based on Josephson junctions are
macroscopic circuits, but they can behave quantum mechanically, like
artificial atoms, allowing the observation of quantum entanglement
and quantum coherence on a macroscopic scale \cite{You-and-Nori-2005,Wendin-and-Shumeiko-2006,Schoelkopf-and-Girvin-2008,Clarke-and-Wilhelm-2008,You-and-Nori-2011,Xiang-et-al-2013,Georgescu-et-al-2014}. These artificial atoms can
be used to test the laws of quantum mechanics on macroscopic
systems and also offer a promising way to implement
quantum information technology.
Superconducting qubits
%play an important role for the
%implementation of
have been widely investigated
theoretically and implemented experimentally in the last fifteen
years due to their advantages, such as scalability, design
flexibility and tunability, for solid-state quantum computation and quantum
simulations \cite{You-and-Nori-2005,Wendin-and-Shumeiko-2006,Schoelkopf-and-Girvin-2008,Clarke-and-Wilhelm-2008,You-and-Nori-2011,Xiang-et-al-2013,Georgescu-et-al-2014,
Pashkin-et-al-2003,Steffen-et-al-2010,Chiorescu-et-al-2004,Wallraff-et-al-2004,Wilson-et-al-2007,Liu-et-al-2005,Valenzuela-et-al-2006,Sillanpää-et-al-2007,Wei-et-al-2008,
Steffen-et-al-2006,Hofheinz-et-al-2009}.
%Superconducting qubits can be controlled by adjusting external
%parameters such as currents, voltages and microwave photons, and
%the coupling between two superconducting qubits can be turned on
%and off.

In practical
applications, the existence of noise (including extrinsic and
intrinsic), inaccuracies (e.g., inaccurate operation in the
coupling between qubits) and fluctuations (e.g., fluctuations in
magnetic fields and electric fields) in superconducting quantum
circuits is unavoidable \cite{McDermott-2009,Valente-et-al-2010PRB,Siddiqi-et-al-2012Nature,Siddiqi-et-al-2012APL,Siddiqi-et-al-2012PRL,Khani-et-al-2012PRA,Paladino-et-al-2014RMP}. For simplicity, in this paper we use fluctuations
to represent noise, inaccuracies, and fluctuations.
These fluctuations degrade the performance of robustness and
reliability in superconducting quantum circuits. Hence, it is
highly desirable, for the development of practical quantum
technology, to design control fields that are robust against
fluctuations \cite{Pravia-et-al-2003,Falci-et-al-2005,Montangero-et-al-2007,Zhang-et-al-2009,Zhang-et-al-2012,Wu-et-al-2013,Kosut-et-al-2013}.

Robustness has been recognized as one of the key properties for a
reliable quantum information processor. Several methods have been
developed for enhancing the robustness of quantum systems \cite{James-et-al-2007,Dong-and-Petersen-2009NJP}. One important
paradigm is feedback control, where the signal obtained from the
system is fed back to adjust input control fields aiming at
achieving improved robustness as well as other measures of system performance
(e.g., stability) \cite{Wiseman-and-Milburn-2009}. A typical
example of the feedback paradigm is quantum error correction, where
possible errors are corrected based on detection outcomes
\cite{Gaitan-2008}. Usually, feedback control is
difficult to implement in practical quantum systems due to the
fast time scale of quantum systems and measurement backaction in
the quantum domain. A more feasible paradigm is open-loop control
for improving robustness of quantum systems where no feedback
signal is required. Dynamical decoupling \cite{Viola-et-al-1999,Khodjasteh-et-al-2010} and optimal control methods \cite{Kosut-et-al-2013,Rabitz-et-al-2004,Wilhelm-et-al-2007,Ginossar-et-al-2010,Wilhelm-et-al-2011,Wilhelm-et-al-2014} can be used
to design robust control fields for manipulating quantum states or
quantum gates. In this paper, we develop a ``smart" open-loop
control method (i.e., sampling-based learning control) to guide
the design of robust control fields for superconducting quantum
systems. The sampling-based learning control (SLC) method includes
two steps of ``training" and ``testing" \cite{Chen-et-al-2013arXiv}. In the training step, we obtain some artificial
samples by sampling the fluctuation parameters and construct an
augmented system using these samples. Then we employ a gradient-flow-based learning algorithm to learn optimal fields for the
augmented system. The robustness of the control fields is tested
and evaluated using additional samples generated by sampling
fluctuation parameters in the testing step. Here we apply the
SLC method to several examples of superconducting
qubits, including single-charge qubits, two coupled charge
qubits and two coupled phase qubits with fluctuations. Our results show that the SLC method
can efficiently learn ``smart" fields that are insensitive to
even 40\%$\sim$50\% fluctuations. The superconducting quantum circuits with
the ``smart" fields can run more reliably.
\newline

\textbf{Results}

\textbf{Single charge qubits with fluctuations.} In
superconducting quantum circuits, the Josephson coupling energy
$E_{J}$ and the charging energy $E_{C}$ are two significant
quantities. Their ratio determines whether the phase or the charge
dominates the behaviour of the qubit, which can form flux qubits
or charge qubits \cite{You-and-Nori-2005}. The simplest charge
qubit is based on a small superconducting island (called a
Cooper-pair box) coupled to the outside world through a weak
Josephson junction and driven by a voltage source through a gate
capacitance within the charge regime (i.e., $E_C\gg E_J$)
\cite{You-and-Nori-2005}. The Hamiltonian of the Cooper-pair box
can be described as \cite{You-and-Nori-2005}
\begin{equation}
H=E_{C}(n-n_{g})^{2}-E_{J}\cos\phi
\end{equation}
where the phase drop $\phi$ across the Josephson junction is
conjugate to the number $n$ of extra Cooper pairs in the box,
$n_{g}=C_{g}V_{g}/2e$ is controlled by the gate voltage $V_{g}$
($C_g$ is the gate capacitance and $2e$ is the charge of each
Cooper pair). In most experiments, the Josephson junction in
the charge qubit is replaced by a dc superconducting quantum interference device (SQUID) to make it easier to control the qubit. In a voltage range near a degeneracy point, the system
can be approximated as a qubit with the following Hamiltonian
\begin{equation}\label{Hamiltonian-1qubit}
H=f(V_g)\sigma_{z}-g(\Phi)\sigma_{x}
\end{equation}
where $f(V_g)$ is related to the charging energy $E_{C}$ and this
term can be adjusted through external parameters, such as the voltage $V_{g}$, and
$g(\Phi)$ corresponds to a controllable term including different
control parameters, such as the flux
$\Phi$ in the SQUID.

For superconducting qubits in laboratories, the existence of
fluctuations is unavoidable (e.g.,
fluctuations in the Josephson coupling energy and the charging
energy, or inaccuracies in the magnetic flux). We assume that possible fluctuations
exist in both $f(V_g)$ and $g(\Phi)$. Suppose that the factors
$f(V_{g})$ and $g(\Phi)$ can be controlled by adjusting
external parameters. Since $E_{J}$ could be around ten GHz and
$E_{C}$ could be around one hundred GHz (e.g., the experiment in
\cite{Pashkin-et-al-2003} used $E_{J1}=13.4\ \text{GHz}$,
$E_{J2}=9.1\ \text{GHz}$, $E_{C1}=117\ \text{GHz}$ and
$E_{C2}=152\ \text{GHz}$), we assume $f(V_{g})/\hbar\in [0,
40]\ \text{GHz}$ and $g(\Phi)/\hbar\in [0, 9.1]\ \text{GHz}$. We could have used 10, instead of 9.1, but we chose 9.1 simply because it was the number used in one experiment. The
practical control terms are $\bar{f}(V_{g})=\theta^{z}f(V_{g})$ and
$\bar{g}(\Phi)=\theta^{x}g(\Phi)$ (where the fluctuation
parameters $\theta^{z}\in [1-\Theta^{z}, 1+\Theta^{z}]$ and
$\theta^{x}\in [1-\Theta^{x}, 1+\Theta^{x}]$) due to possible
multiplicative fluctuations. Here the bounds of fluctuations
$\Theta^{z}$ and $\Theta^{x}$ characterize the maximum ranges of fluctuations in $\theta^{z}$ and $\theta^{x}$, respectively. The fluctuations can originate from the fluctuations in the magnetic flux $\Phi$, the
voltage $V_{g}$, the Josephson coupling energy $E_{J}$ and the
charging energy $E_{C}$.

\begin{figure}
\center{\includegraphics[scale=0.65]{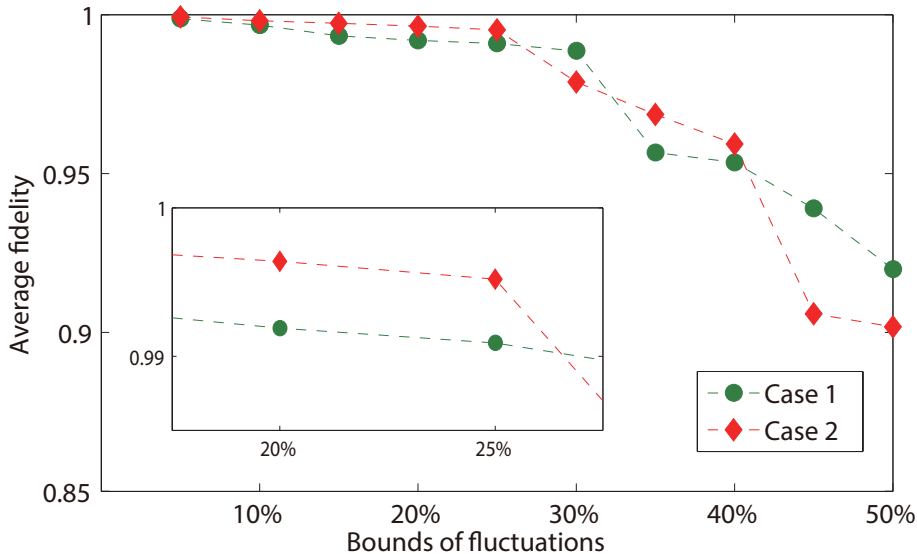}}% Here is how to import EPS art
\caption{\label{fig1} Average
fidelity versus the bounds of fluctuations $\Theta^{z}$ and
$\Theta^{x}$ for charge qubits. The fluctuation parameters $\theta^{z}$ and
$\theta^{x}$ have uniform distributions in $[1-\Theta,
1+\Theta]$ (i.e., we assume $\Theta^{z}=\Theta^{x}=\Theta$).
Here we consider $|\psi_{0}\rangle=|g\rangle$, and
$|\psi_{\text{target}}\rangle=|e\rangle$ (Case 1) or
$|\psi_{\text{target}}\rangle=\frac{1}{\sqrt{2}}(|g\rangle+|e\rangle)$
(Case 2). Every average fidelity is calculated using 5000 tested
samples.}
\end{figure}

As an example, we assume that the initial state is
$|\psi_{0}\rangle=|g\rangle$, and the target state is either
$|\psi_{\text{target}}\rangle=|e\rangle$ or
$|\psi_{\text{target}}\rangle=\frac{1}{\sqrt{2}}(|g\rangle+|e\rangle)$.
Let the operation time be $T=1\ \text{ns}$. Now we employ the
sampling-based learning control method (see the methods Section)
to learn an optimal control field for manipulating the charge
qubit system from the initial state to a target state. The time
interval $t\in[0, 1]\ \text{ns}$ is equally divided into $100$
smaller time intervals. Without loss of generality, we assume $\theta^{x}$ and $\theta^{z}$ to have
uniform distributions and have the same bound of fluctuations (i.e., $\Theta^{x}=\Theta^{z}$). An augmented
system is constructed by selecting $N_{x}=5$ values for
$\theta^{x}$ and $N_{z}=5$ values for $\theta^{z}$. The initial
control fields are $f(V_g)/\hbar=\sin t+\cos t+20$ GHz and
$g(\Phi)/\hbar=\sin t+\cos t+5$ GHz. The learning algorithm runs for about
$7000$ iterations for $|\psi_{\text{target}}\rangle=|e\rangle$
($4000$ iterations for
$|\psi_{\text{target}}\rangle=\frac{1}{\sqrt{2}}(|g\rangle+|e\rangle)$)
before it converges to optimal control fields.
After the optimal control fields are learned for the augmented
system, they are applied to 5000 samples generated by
stochastically selecting the values of the fluctuation
parameters for evaluating the performance. The fidelity between
the final state $|\psi(T)\rangle$ and the target state
$|\psi_{\text{target}}\rangle$ is defined as
$F(|\psi(T)\rangle,|\psi_{\text{target}}\rangle)=|\langle
\psi(T)|\psi_{\text{target}}\rangle|$ \cite{Nielsen-and-Chuang-2000}. The relationship between the average fidelity and the
bounds of the fluctuations is shown in Fig. \ref{fig1}. Although
the performance decreases when increasing the bounds on the
fluctuations, the ``smart" fields can still drive the system from
the initial state $|\psi_{0}\rangle=|g\rangle$ to a given target
state with high fidelity (the average fidelity is $\bar{F}=0.9909$
for $|\psi_{\text{target}}\rangle=|e\rangle$, and $\bar{F}=0.9952$
for
$|\psi_{\text{target}}\rangle=\frac{1}{\sqrt{2}}(|g\rangle+|e\rangle)$)
even though the bound on the fluctuations is $25\%$ (i.e., $50\%$ fluctuations relative to the nominal value).

\begin{figure}
\center{\includegraphics[scale=0.65]{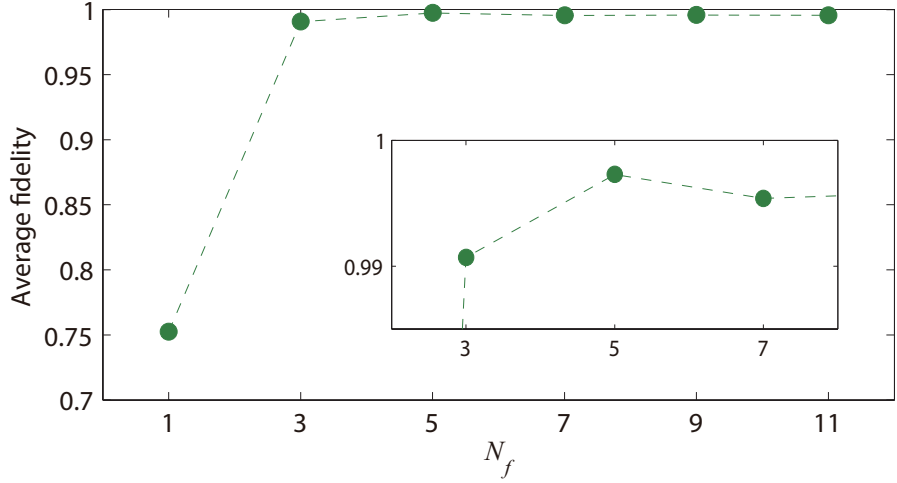}}% Here is how to import EPS art
\caption{\label{fig2} Average fidelity versus the number $N_{f}$ of values for $\theta^{x}$ and $\theta^{z}$ ($N_{f}=N_{x}=N_{z}$). Here,
$\Theta^{z}=\Theta^{x}=15\%$ (i.e., 30\% fluctuations),
$|\psi_{0}\rangle=|g\rangle$ and
$|\psi_{\text{target}}\rangle=\frac{1}{\sqrt{2}}(|g\rangle+|e\rangle)$.
Every average fidelity is calculated using 5000 samples.}
\end{figure}

We also test the relationship between the number of values $N_{f}$ for $\theta^{x}$ and $\theta^{z}$
($N_x=N_z=N_{f}$) and the average fidelity for bounds on the
fluctuations $\Theta^{z}=\Theta^{x}=15\%$. The performance is
shown in Fig. \ref{fig2}. It is clear that the performance is
excellent for $N_{f}=5$ or $7$. Although it is possible to improve the performance through using more samples,
too many samples will cost more
computation resources and spend too much time for learning a set of
optimal fields. For example, the laptop Thinkpad T440p, with a CPU of 2.5 GHz, needs about 13 minutes to find the optimal fields for $N_f=5$; while this laptop requires about 42 minutes for $N_f=11$. When increasing the number of fluctuation parameters, the time consuming quickly increases with the increasing of $N_{f}$. Hence, we choose $N_f=5$
for each fluctuation parameter in all of the following numerical results.
\newline

\textbf{Two coupled qubits with fluctuations.}

\begin{figure}
\center{\includegraphics[scale=0.2]{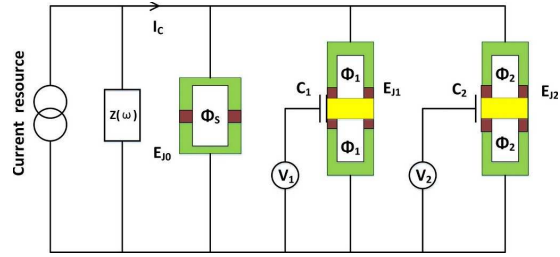}}% Here is how to import EPS art
\caption{\label{fig3} A coupled-qubit circuit with a
biased-current source of impedance $Z(\omega)$ \cite{You-et-al-2003}. Two charge qubits
are coupled by the dc SQUID with two junctions with large $E_{J0}$.}
\end{figure}

We first consider the coupled qubit circuit in \cite{You-et-al-2003}
where a symmetric dc SQUID with two sufficiently large junctions
is used to couple two charge qubits (see Fig. \ref{fig3}). Each
qubit is realized by a Cooper-pair box with Josephson coupling
energy $E_{Jj}$ and capacitance $C_{Jj}$ (j=1, 2). Each
Cooper-pair box is biased by an applied voltage $V_{j}$ through
the gate capacitance $C_{j}$ ($j=1, 2$). We apply a flux
$\Phi_{s}$ inside the large-junction dc SQUID loop with two
junctions of large $E_{J0}$. The Hamiltonian of the coupled charge
qubits can be described as
\begin{equation}
H=f(V_{1})\sigma^{(1)}_{z}+f(V_{2})\sigma^{(2)}_{z}-g(\Phi_1)\sigma^{(1)}_{x}-g(\Phi_2)\sigma^{(2)}_{x}-\chi(t)\sigma^{(1)}_{x}\sigma^{(2)}_{x}.
\end{equation}
Due to possible fluctuations, we assume that the Hamiltonian for practical systems is
 \begin{equation}\label{coupledHamiltonian1}
H=\theta_{1}f(V_{1})\sigma^{(1)}_{z}+\theta_{2}f(V_{2})\sigma^{(2)}_{z}-\theta_{3}g(\Phi_1)\sigma^{(1)}_{x}-\theta_{4}g(\Phi_2)\sigma^{(2)}_{x}-\theta_{5}\chi(t)\sigma^{(1)}_{x}\sigma^{(2)}_{x}
\end{equation}
where the fluctuation parameters $\theta_{j}\in [1-\Theta_{j}, 1+\Theta_{j}]$ ($j=1,2,3,4,5$).

\begin{figure}
\center{\includegraphics[scale=0.65]{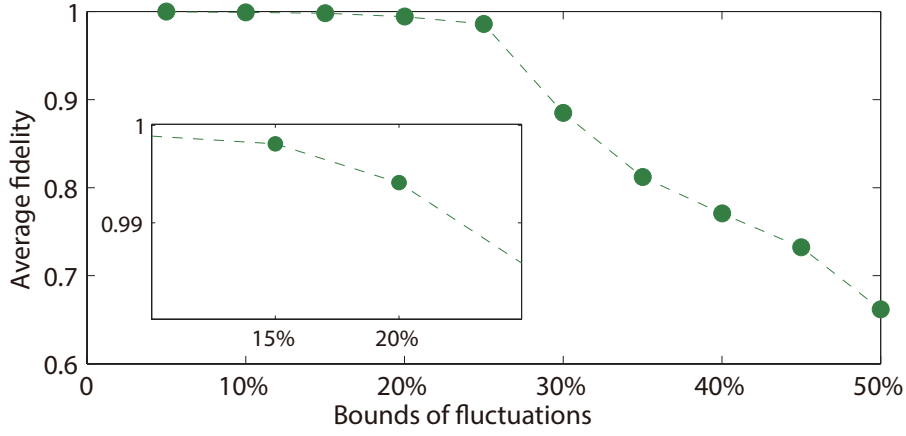}}% Here is how to import EPS art
\caption{\label{fig4} Average
fidelity versus the bound on the fluctuations $\Theta$ for two coupled
charge qubits with a biased-current source. The fluctuation
parameters $\theta_{j}=1-\vartheta_{j}\cos t$ ($j=1, 2, 3, 4$),
where each $\vartheta_{j}$ has a uniform distribution in $[1-\Theta,
1+\Theta]$. Here we assume $\theta_{1}=\theta_{2}$ and
$\theta_{3}=\theta_{4}$. The initial state
$|\psi_{0}\rangle=0.7|g,g\rangle+0.1|g,e\rangle+0.7i|e,g\rangle+0.1i|e,e\rangle$
and the target state
$|\psi_{\text{target}}\rangle=0.7|g,g\rangle+0.1|g,e\rangle+0.7i|e,g\rangle-0.1i|e,e\rangle$.
Each average fidelity is calculated using 5000 samples. }
\end{figure}

We let $g(\Phi_1)/\hbar=g(\Phi_2)/\hbar=9.1\ \text{GHz}$, the
control terms $f(V_1)/\hbar \in [0, 40]\ \text{GHz}$, $f(V_2)/\hbar
\in [0, 40]\ \text{GHz}$, $|\chi(t)/\hbar |\leq 0.5\ \text{GHz}$ and
$\theta_{5}(t)\equiv1$. The operation time $T=2\ \text{ns}$. We assume that the
fluctuation parameters $\theta_{j}$ ($j=1, 2, 3, 4$) may be time
varying. Hence, $\theta_{3}$ and $\theta_{4}$ may correspond to
time-varying additive fluctuations. The fluctuations in $\theta_1$ and $\theta_2$ may originate
from the time-varying errors in the driving fields. As an illustrative example, we
let $\theta_{j}=1-\vartheta_{j}\cos t$, where each $\vartheta_{j}$ has
a uniform distribution in the interval $[1-\Theta_{j},
1+\Theta_{j}]$. For simplification, we assume
$\theta_{1}=\theta_{2}$, $\theta_{3}=\theta_{4}$ and
$\Theta_{1}=\Theta_{2}=\Theta_{3}=\Theta_{4}=\Theta$. We now consider a controlled-phase-shift gate operation on an initial
state
$|\psi_{0}\rangle=\alpha_{1}|g,g\rangle+\alpha_{2}|g,e\rangle+\alpha_{3}|e,g\rangle+\alpha_{4}|e,e\rangle$;
i.e., the target state is
$|\psi_{\text{target}}\rangle=\alpha_{1}|g,g\rangle+\alpha_{2}|g,e\rangle+\alpha_{3}|e,g\rangle-\alpha_{4}|e,e\rangle$.
In particular, we let $\alpha_{1}=0.7$, $\alpha_{2}=0.1$,
$\alpha_{3}=0.7i$ and $\alpha_{4}=0.1i$. The time interval $t\in
[0, 2]\ \text{ns}$ is equally divided into $200$ smaller time
intervals. The control fields are initialized as:
$f(V_1)/\hbar=f(V_2)/\hbar=\sin t+\cos t+5$ GHz,
$\chi(t)/\hbar=0.25\sin t$ GHz. The learning algorithm runs for
about $5000$ iterations before the optimal control fields are found.
Then the learned fields are applied to 5000 samples that are
generated by selecting the values of
the fluctuation parameters according to a uniform distribution. The performance is shown in Fig.
\ref{fig4}. Although the performance decreases when
increasing the bounds on the fluctuations, the ``smart" fields can still
drive the system from the initial state
$|\psi_{0}\rangle=0.7|g,g\rangle+0.1|g,e\rangle+0.7i|e,g\rangle+0.1i|e,e\rangle$
to the target state
$|\psi_{\text{target}}\rangle=0.7|g,g\rangle+0.1|g,e\rangle+0.7i|e,g\rangle-0.1i|e,e\rangle$
with high fidelity (average fidelity 0.9941) \emph{even with} $40\%$
\emph{fluctuations}.

In the two numerical examples of single-charge qubits and two coupled charge qubits, we use some ideal parameter values to show the
effectiveness and excellent performance of the proposed method. It is straightforward to extend our method to other systems. Indeed, our proposed method is very flexible in the selection of the operation time $T$ and the target state, and is also robust against fluctuations with different distributions. Here, we consider another example based on the two coupled phase qubits in Ref. \cite{Martinis-et-al-2011PRL}. Each phase qubit is a nonlinear resonator built from an Al/AlO$_x$/Al Josephson junction, and two qubits are coupled via a modular four-terminal device (for detail, see Fig. 1 in \cite{Martinis-et-al-2011PRL}). This four-terminal device is constructed using two nontunable inductors, a fixed mutual inductance and a tunable inductance. The equivalent Hamiltonian can be described as
\begin{equation}
H=\frac{\hbar \omega_{1}(t)}{2}\sigma^{(1)}_{z}+\frac{\hbar \omega_{2}(t)}{2}\sigma^{(2)}_{z}+\frac{\hbar \omega_{3}(t)}{2}\sigma^{(1)}_{x}+\frac{\hbar \omega_{4}(t)}{2}\sigma^{(2)}_{x}+\frac{\hbar \Omega_{c}(t)}{2}(\sigma^{(1)}_{x}\sigma^{(2)}_{x}+\frac{1}{6\sqrt{N_{1}N_{2}}}\sigma^{(1)}_{z}\sigma^{(2)}_{z})
\end{equation}
where $N_{1}$ and $N_{2}$ are the number of levels in the potentials of qubits 1 and 2. The typical values for $N_{1}$ and $N_{2}$ are $N_{1}=N_{2}=5$.
Due to possible fluctuations, we assume that the practical Hamiltonian has the following form
\begin{equation}\label{coupledHamiltonian2}
H=\frac{\hbar \theta_{1} \omega_{1}(t)}{2}\sigma^{(1)}_{z}+\frac{\hbar \theta_{2} \omega_{2}(t)}{2}\sigma^{(2)}_{z}+\frac{\hbar \omega_{3}}{2}\sigma^{(1)}_{x}+\frac{\hbar \omega_{4}}{2}\sigma^{(2)}_{x}+\frac{\hbar \theta_{3} \Omega_{c}(t)}{2}(\sigma^{(1)}_{x}\sigma^{(2)}_{x}+\frac{1}{30}\sigma^{(1)}_{z}\sigma^{(2)}_{z})
\end{equation}
with $\theta_{j}\in [1-\Theta, 1+\Theta]$ ($j=1,2,3$).

We assume that the frequencies $\omega_{1}(t), \omega_{2}(t) \in [0, 5]\ \text{GHz}$ can be adjusted by changing the bias currents of two phase qubits, and $\Omega_{c}(t) \in [-100, 100]\ \text{MHz}$ can be adjusted by changing the bias current in the coupler. Let $\omega_{3}=\omega_{4}=2\ \text{GHz}$, the operation time $T=50 \ \text{ns}$, and each
fluctuation parameter $\theta_{j}$ $(j=1,2,3)$ in
(\ref{coupledHamiltonian2}) has a truncated Gaussian distribution in
$[1-\Theta, 1+\Theta]$. Assume that the probability density function
of the truncated Gaussian distribution is
$p(x,\mu,\sigma,l,r)=\phi(\frac{x-\mu}{\sigma})\{\sigma[\Phi(\frac{r-\mu}{\sigma})
-\Phi(\frac{l-\mu}{\sigma})]\}^{-1}$, where $\mu=0$,
$\sigma=\Theta/3$, $l=-\Theta$, $r=\Theta$, $\phi(x)={(2\pi)}^{-1/2}\text{exp}(-\frac{1}{2}x^2)$
is the probability density function of the standard normal
distribution, and $\Phi(x)$ is its cumulative distribution function.

\begin{figure}
\center{\includegraphics[scale=0.65]{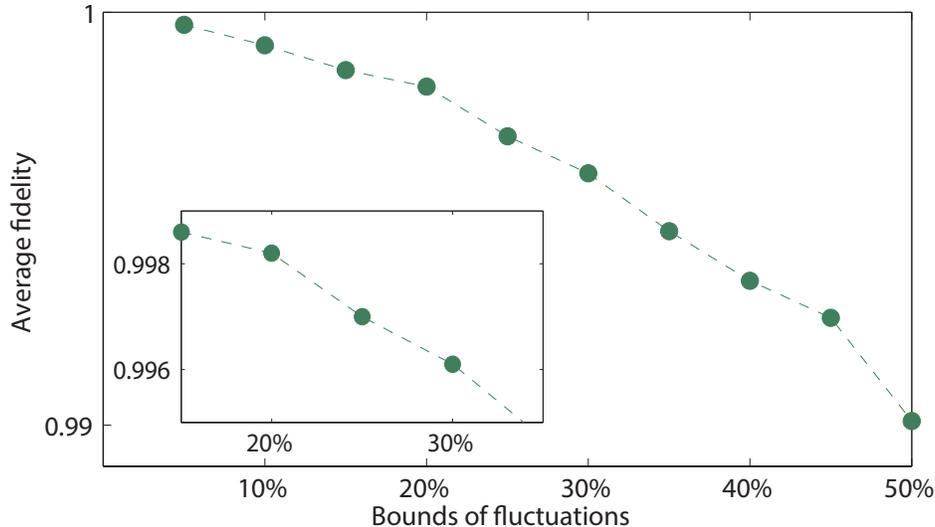}}% Here is how to import EPS art
\caption{\label{fig5} Average
fidelity versus the bound on the fluctuations $\Theta$ for two coupled
phase qubits. Each $\theta_{j}$
($j=1, 2, 3$) has a truncated Gaussian distribution in
$[1-\Theta, 1+\Theta]$, and we assume $\theta_{1}=\theta_{2}$. The initial state is
$|\psi_{0}\rangle=\frac{1}{\sqrt{2}}(|g,g\rangle+|e,g\rangle)$ and
the target state is
$|\psi_{\text{target}}\rangle=\frac{1}{\sqrt{2}}(|g,g\rangle+|e,e\rangle)$.
Each average fidelity is calculated using 5000 samples. }
\end{figure}

\begin{figure}
\center{\includegraphics[scale=0.75]{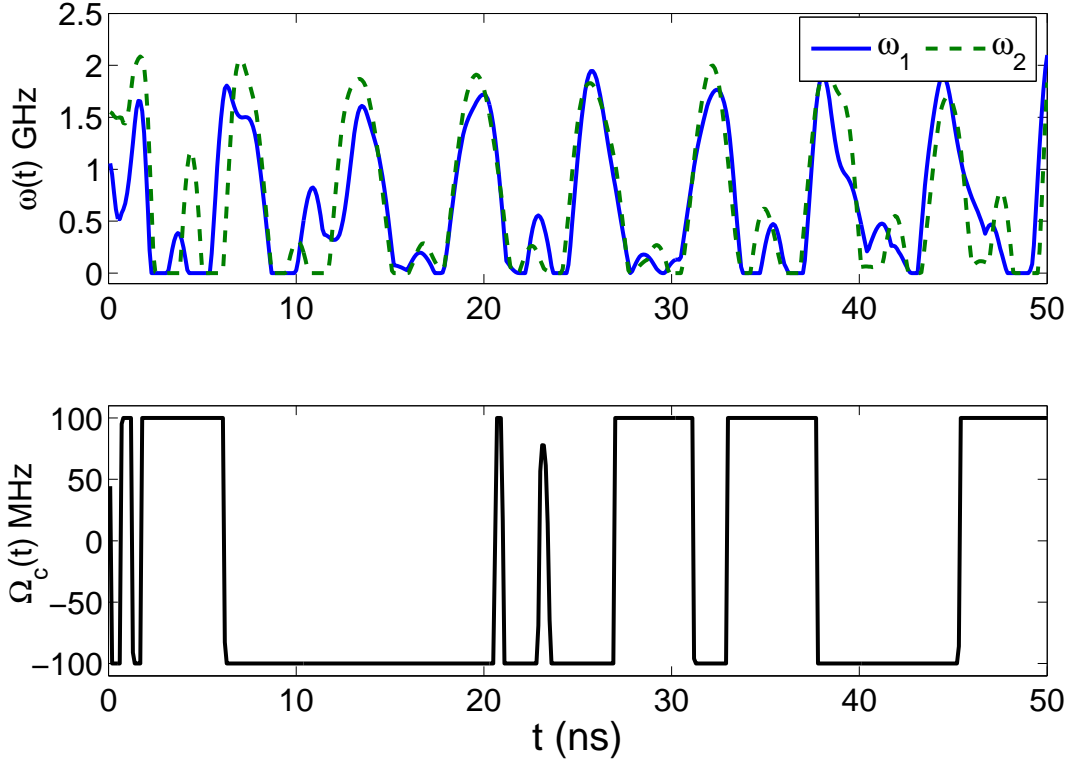}}% Here is how to import EPS art
\caption{\label{fig6} A set of ``smart" fields corresponding to $\omega_{1}(t)$, $\omega_{2}(t)$ and $\Omega_{c}(t)$ for the problem of
two coupled phase qubits when the bound
on the fluctuations is very large, with $\Theta=25\%$ (i.e., 50\% fluctuations). }
\end{figure}

We now consider a CNOT operation. In
particular, we let the initial state be
$|\psi_{0}\rangle=\frac{1}{\sqrt{2}}(|g,g\rangle+|e,g\rangle)$ and
the target state be a maximum entangled state
$|\psi_{\text{target}}\rangle=\frac{1}{\sqrt{2}}(|g,g\rangle+|e,e\rangle)$. In the training step, the
fluctuations are uniformly sampled. However, in the testing step the
samples are selected by sampling the fluctuation parameters with a truncated Gaussian
distribution. For simplicity, we let $\theta_{1}=\theta_{2}$. The initial vaules are $\omega_{1}=\omega_{2}=\sin t+\cos t+0.5$ GHz and $\Omega_{c}(t)=50+50\sin t$ MHz. Other parameter settings are the same as those in the example of coupled charge qubits. The performance is shown in Fig.
\ref{fig5} and a
set of ``smart" fields is shown in Fig. \ref{fig6} for
$\Theta=25\%$ (i.e., 50\% fluctuations). The ``smart" fields can drive the system from $|\psi_{0}\rangle$ to $|\psi_{\text{target}}\rangle$ with high fidelity (average fidelity 0.9970) even with 50\% fluctuations.
\newline

\textbf{Discussion}

In numerical examples, a small number of samples for each possible fluctuation parameter is used to construct an augmented system. It is possible to achieve improved performance by using more samples. However, an increase in the number of samples in the training step will consume more computation resources. The tradeoff between  resource consumption and performance that can be achieved should be investigated for specific tasks. In the SLC method, we employ a general gradient-flow-based algorithm to learn ``smart" fields and the algorithm is usually much more efficient than other stochastic search algorithms (e.g., genetic algorithms) for control design of quantum systems \cite{Rabitz-et-al-2004}. The ``smart" fields are ``optimal" to the control landscapes of different samples since they are found by simultaneously optimizing the fields for these samples. It may be possible to use a similar theory to the quantum control landscape theory developed in \cite{Rabitz-et-al-2004} to analyze these optimal properties. As examples, we only consider that each possible fluctuation parameter has several specific distributions in the testing step. However, the proposed method also works well for other time-varying or time-invariant distributions. Numerical results show that, in the training step, sampling fluctuation parameters according to simple uniform distributions can achieve excellent performance for these cases where the fluctuation parameters have other distributions. In our numerical examples, we only consider three classes of superconducting quantum systems with several specific parameter settings. Our method is also applicable to other superconducting qubits, such as the ``Xmon" and ``gmon" qubits \cite{Martinis-et-al-2013PRL,Martinis-et-al-2014-1,Martinis-et-al-2014-2}, since its performance is insensitive to the parameter settings and possible fluctuations.

In conclusion, we use a sampling-based learning control method to design robust fields that are \emph{insensitive} to possible fluctuations. Numerical results show that the method can efficiently find ``smart" fields for superconducting qubits \emph{even in the presence of 40\%$\sim$50\% fluctuations} in different parameters. The proposed method has potential for robust quantum information processing and can contribute to the design of more reliable quantum devices.
\newline

\textbf{Methods}

\textbf{Sampling-based learning control (SLC).} The SLC method was first proposed for
the control design of inhomogeneous quantum ensembles \cite{Chen-et-al-2013arXiv}. In this method, several artificial samples, generated
through sampling possible inhomogeneous parameters, are used to learn
optimal control fields and then these fields are applied to
additional samples to test their performance. In this paper, we develop an SLC
method for guiding the design of robust control fields for superconducting
quantum systems with fluctuations.

Consider a quantum system with Hamiltonian $H(u,\theta, t)$ and the
evolution of its state $|\psi(t)\rangle$ is described by the
following Schr\"{o}dinger equation:
\begin{equation} \label{systemmodel}
  i \hbar|{\dot{\psi}}(t)\rangle=H(u,\theta, t)|\psi(t)\rangle
\end{equation}
where $u$ represents the control field and $\theta$ characterizes
possible fluctuations. In the SLC method, we first generate $N$ artificial
samples by selecting different values of $\theta$ (e.g., the $N$ samples
correspond to $\theta^{(1)}$,
$\theta^{(2)}$,$\cdots$, $\theta^{(N)}$). Using these samples, an augmented system is constructed as follows
\begin{equation}\label{augmented-system}
i\hbar\left(%
\begin{array}{c}
  |{\dot{\psi}}_{\theta^{(1)}}(t)\rangle \\
  |{\dot{\psi}}_{\theta^{(2)}}(t)\rangle \\
  \vdots \\
  |{\dot{\psi}}_{\theta^{(N)}}(t)\rangle \\
\end{array}%
\right)
=\left(%
\begin{array}{c}
  H(u,\theta^{(1)}, t)|\psi_{\theta^{(1)}}(t)\rangle \\
  H(u,\theta^{(2)}, t)|\psi_{\theta^{(2)}}(t)\rangle \\
  \vdots \\
  H(u,\theta^{(N)}, t)|\psi_{\theta^{(N)}}(t)\rangle \\
\end{array}%
\right).
\end{equation}
The performance function $J(u)$ for the augmented
system is defined as
\begin{equation}\label{eq:cost}
J(u):=\frac{1}{N}\sum_{n=1}^N J_{\theta^{(n)}}(u)=\frac{1}{N}\sum_{n=1}^{N}\vert \langle\psi_{\theta^{(n)}}(T)|\psi_{\text{target}}\rangle\vert^{2}
\end{equation}
where $|\psi_{\text{target}}\rangle$ is the target state and
$|\psi_{\theta^{(n)}}(T)\rangle$ is the final state for one sample
(corresponding to $\theta^{(n)}$) at the time $T$. The task in the
training step is to find an optimal control field $u^*$ that
maximizes the performance function defined in Eq. \eqref{eq:cost}.

In the testing step, we apply the optimized field $u^{*}$ to
additional samples generated by randomly sampling the
fluctuation parameters and evaluate the performance in terms of the
fidelity. If the average fidelity for the tested samples are good
enough, we accept the designed field and complete the design
process. Otherwise, we should go back to the training step and
learn another optimized control field (e.g., restarting the
training step with a new initial field or a new set of samples).
\newline

\textbf{Sampling.}
In order to construct an augmented system, we
need to generate $N$ artificial samples. We assume that there are two fluctuation
parameters $\theta^{x}$ and $\theta^{z}$. We may choose some
equally-spaced samples in the $\theta^{x}$--$\theta^{z}$ space. For
example, the intervals $[1-\Theta^{x}, 1+\Theta^{x}]$ and
$[1-\Theta^{z}, 1+\Theta^{z}]$ are divided into $N_{x}+1$ and
$N_{z}+1$ subintervals, respectively, where $N_{x}$ and $N_{z}$
are usually positive odd numbers. Then, the number of samples
$N=N_{x}N_{z}$, where $\theta^{x}_{m}$ and $\theta^{z}_{n}$ can be
chosen from the combination of $(\theta^{x}_{m}, \theta^{z}_{n})$
as follows
\begin{equation}\label{discrete}
\left\{ \begin{array}{c} \theta^{x}_{m} \in
\{\theta^{x}_{m}=1-\Theta^{x}+\frac{(2m-1)\Theta^{x}}{N_{x}},
\ m=1,2,\ldots, N_{x}\},\\
\theta^{z}_{n} \in
\{\theta^{z}_{n}=1-\Theta^{z}+\frac{(2n-1)\Theta^{z}}{N_{z}},\
\
n=1,2,\ldots, N_{z}\}. \\
\end{array}
\right.
\end{equation}
\newline

\textbf{Gradient-flow-based learning algorithm.}
In order to find an optimal control field $u^{*}$ for the
augmented system (\ref{augmented-system}), a good choice is to
follow the direction of the gradient of $J(u)$ as an ascent
direction. Assume that the performance function is $J(u_{0})$ with
an initial field $u_{0}$. We can apply the gradient flow
method to approximate an optimal control field $u^{*}$. This can
be achieved by iterative learning using the following updating (for
details, see, e.g., \cite{Chen-et-al-2013arXiv})
\begin{equation}\label{iteration2}
u_{k+1}(t)=u_{k}(t)+ \eta_{k}\nabla J(u_{k}),
\end{equation}
where $\eta_{k}$ is the updating stepsize for the $k$th iteration
and $\nabla J(u)$ denotes the gradient of $J(u)$ with respect to
the control $u$. The calculation of $\nabla J(u)$ is described in
\cite{Chen-et-al-2013arXiv,Roslund-and-Rabitz-2009}. For
practical implementations, we usually divide the time interval
$[0,T]$ equally into a number of smaller time intervals $\Delta t$
and assume that the control fields are constant within each time
interval $\Delta t$. In the algorithm, we assume $u(t)\in [V_{-}, V_{+}]$. If
$u_{k+1}\leq V_{-}$, we let $u_{k+1}= V_{-}$. If $u_{k+1}\geq
V_{+}$, we let $u_{k+1}= V_{+}$. In numerical computations, if the change of the performance function
for $100$ consecutive training steps is less than a small threshold $\epsilon$ (i.e., $|J(u_{k+100})-J(u_k)|<\epsilon$ for some $k$), then
the algorithm converges and we end the training step.
In this paper, we let $\epsilon=10^{-4}$ for all numerical results.

\textbf{Acknowledgments}
The work is supported by the Australian Research Council (DP130101658, FL110100020) and National Natural Science Foundation of China (Nos. 61273327 and 61374092). F.N. is partially supported by the RIKEN iTHES Project, MURI Center for Dynamic Magneto-Optics, and a Grant-in-Aid for Scientific Research (S).
\newline

\textbf{Author contributions}
D.D., C.C., B.Q. and I.R.P. developed the scheme of robust sampling-based learning control, D.D. and F.N. designed the illustrative examples of superconducting quantum circuits. C.C. performed the numerical simulations.
All authors discussed the results and contributed to the writing of the paper.
\newline

\textbf{Additional information}

\textbf{Competing financial interests:} The authors declare no competing financial interests.


\begin{thebibliography}{99}

\bibitem{You-and-Nori-2005}
You, J. Q. \& Nori, F. Superconducting circuits and quantum information. {\em Physics Today}
\textbf{58}, 42-47 (2005).

\bibitem{Wendin-and-Shumeiko-2006}
Wendin, G. \& Shumeiko, V. S. In \emph{Handbook of Theoretical and Computational Nanotechnology}, edited by M. Rieth and W. Schommers (American Scientific Publishers, Karlsruhe, Germany, 2006), Chap. 12; arXiv:cond-mat/0508729

\bibitem{Schoelkopf-and-Girvin-2008}
Schoelkopf, R. J. \& Girvin, S. M. Wiring up quantum systems. {\em Nature}
\textbf{451}, 664-669 (2008).

\bibitem{Clarke-and-Wilhelm-2008}
Clarke, J. \& Wilhelm, F. K. Superconducting quantum bits. {\em Nature}
\textbf{453}, 1031-1042 (2008).

\bibitem{You-and-Nori-2011}
You, J. Q. \& Nori, F. Atomic physics and quantum optics using superconducting circuits. {\em Nature}
\textbf{474}, 589-597 (2011).

\bibitem{Xiang-et-al-2013}
Xiang, Z. L., Ashhab, S., You, J. Q. \& Nori, F. Hybrid quantum circuits: Superconducting circuits interacting with other quantum systems. {\em Rev. Mod. Phys.} \textbf{85}, 623-653 (2013).

%\bibitem{Makhlin et al 2001}
%Y. Makhlin, G. Sch\"{o}n and A. Shnirman, ``Quantum-state engineering with Josephson-junction devices," {\em Reviews of Modern Physics},
%Vol.73, 357-400, 2001.

%\bibitem{Yamamoto et al 2003}
%T. Yamamoto, Yu. A. Pashkin, O. Astafiev, Y. Nakamura and J. S. Tsai, ``Demonstration of conditional gate operation using superconducting charge qubits," {\em Nature},
%Vol.425, 941-944, 2003.

\bibitem{Georgescu-et-al-2014}
Georgescu, I., Ashhab, S. \& Nori, F. Quantum Simulation. {\em Rev. Mod. Phys.} \textbf{86}, 153-185 (2014).

\bibitem{Pashkin-et-al-2003}
Pashkin, Yu. A., \emph{et al.} Quantum oscillations in two coupled charge qubits. {\em Nature} \textbf{421}, 823-826 (2003).

\bibitem{Steffen-et-al-2010}
Steffen, M., \emph{et al.} High-coherence hybrid superconducting qubit. {\em Phys. Rev. Lett.}
\textbf{105}, 100502 (2010).

\bibitem{Chiorescu-et-al-2004}
Chiorescu, I., \emph{et al.} Coherent dynamics of a flux qubit coupled to a harmonic oscillator. {\em Nature} \textbf{431}, 159-162 (2004).

\bibitem{Wallraff-et-al-2004}
Wallraff, A., \emph{et al.} Strong coupling of a single photon to a superconducting qubit using circuit quantum electrodynamics. {\em Nature} \textbf{431}, 162-167 (2004).

\bibitem{Wilson-et-al-2007}
Wilson, C. M., \emph{et al.} Observation of the dynamical Casimir effect in a superconducting circuit. {\em Nature} \textbf{479}, 376-379 (2011).
%Wilson, C. M., Duty, T., Persson, F., Sandberg, M., Johansson, G. \& Delsing, P. Coherence times of dressed states of a superconducting qubit under extreme driving. {\em Phys. Rev. Lett.} %\textbf{98}, 257003 (2007).

\bibitem{Liu-et-al-2005}
Liu, Y. X., You, J. Q.,Wei, L. F., Sun, C. P. \& Nori, F. Optical selection rules and phase-dependent adiabatic state control in a superconducting quantum circuit. {\em Phys. Rev. Lett.} \textbf{95}, 087001 (2005).

\bibitem{Valenzuela-et-al-2006}
Valenzuela, S.O., \emph{et al.} Microwave-induced cooling of a superconducting qubit. {\em Science} \textbf{314}, 1589-1592 (2006).

\bibitem{Sillanpää-et-al-2007}
Sillanp\"{a}\"{a}, M. A., Park, J. I. \& Simmonds, R. W. Coherent quantum state storage and transfer between two phase qubits via a resonant cavity. {\em Nature} \textbf{449}, 438-442 (2007).

\bibitem{Wei-et-al-2008}
Wei, L. F., Johansson, J. R., Cen, L. X., Ashhab, S. \& Nori, F. Controllable coherent population transfers in superconducting qubits for quantum computing. {\em Phys. Rev. Lett.} \textbf{100}, 113601 (2008).

\bibitem{Steffen-et-al-2006}
Steffen, M., \emph{et al.} Measurement of the entanglement of two superconducting qubits via state tomography. {\em Science} \textbf{313}, 1423-1425 (2006).

\bibitem{Hofheinz-et-al-2009}
Hofheinz, M., \emph{et al.} Synthesizing arbitrary quantum states in a superconducting resonator. {\em Nature} \textbf{459}, 546-549 (2009).

\bibitem{McDermott-2009}
McDermott, R. Materials orignins of decoherence in superconducting qubits.
{\em IEEE Trans. Appl. Superconductivity.} \textbf{19}, 2-13 (2009).

\bibitem{Valente-et-al-2010PRB}
Valente, D. C. B., Mucciolo, E. R. \& Wilhelm, F. K. Decoherence by electromagnetic fluctuations in double-quantum-dot charge qubits.
{\em Phys. Rev. B}, \textbf{82}, 125302 (2010).

\bibitem{Siddiqi-et-al-2012Nature}
Vijay, R., \emph{et al.} Stabilizing Rabi oscillations in a superconducting qubit using quantum feedback. {\em Nature} \textbf{490}, 77-80 (2012).

\bibitem{Siddiqi-et-al-2012APL}
Murch, K. W., Weber, S. J., Levenson-Falk, E. M., Vijay, R. \& Siddiqi, I. $1/f$ noise of Josephson-junction-embedded microwave resonators at single photon energies and millikelvin temperatures. {\em Appl. Phys. Lett.} \textbf{100}, 142601 (2012).

\bibitem{Siddiqi-et-al-2012PRL}
Slichter, D. H., \emph{et al.} Measurement-induced qubit state mixing in circuit QED from up-converted dephasing noise.
{\em Phys. Rev. Lett.} \textbf{109}, 153601 (2012).

\bibitem{Khani-et-al-2012PRA}
Khani, B., Merkel, S. T., Motzoi, F., Gambetta, J. M. \& Wilhelm, F. K. High-fidelity quantum gates in the presence of dispersion.
{\em Phys. Rev. A} \textbf{85}, 022306 (2012).

\bibitem{Paladino-et-al-2014RMP}
Paladino, E., Galperin, Y. M., Falci, G. \& Altshuler, B. L. $1/f$ noise: Implications for solid-state quantum information.
{\em Rev. Mod. Phys.} \textbf{86}, 361-418 (2014).


\bibitem{Pravia-et-al-2003}
Pravia, M. A.,\emph{ et al.} Robust control of quantum information.
{\em J. Chem. Phys.} \textbf{119}, 9993-10001 (2003)

\bibitem{Falci-et-al-2005}
Falci, G., D'Arrigo, A., Mastellone, A. \& Paladino, E. Initial decoherence in solid state qubits. {\em Phys. Rev. Lett.} \textbf{94}, 167002 (2005).

\bibitem{Montangero-et-al-2007}
Montangero, S., Calarco, T. \& Fazio, R. Robust optimal quantum gates for Josephson charge qubits. {\em Phys. Rev. Lett.} \textbf{99}, 170501 (2007).

\bibitem{Zhang-et-al-2009}
Zhang, J., Liu, Y. X. \& Nori, F. Cooling and squeezing the fluctuations of a nanomechanical beam by indirect quantum feedback control. {\em Phys. Rev. A} \textbf{79}, 052102 (2009).

\bibitem{Zhang-et-al-2012}
Zhang, J., Greenman, L., Deng, X. \& Whaley, K. B. Robust control pulses
design for electron shuttling in solid state devices. {\em IEEE Trans. Control Syst. Technology} \textbf{22}, 2354-2359 (2014).

\bibitem{Wu-et-al-2013}
Wu, R. B., \emph{et al.} Spectral analysis and identification of noises in quantum systems. {\em Phys. Rev. A} \textbf{87}, 022324 (2013).

\bibitem{Kosut-et-al-2013}
Kosut, R. L., Grace, M. D. \& Brif, C. Robust control of quantum gates via sequential convex programming. {\em Phys. Rev. A} \textbf{88}, 052326 (2013).

%\bibitem{James 2004}
%M.R. James, ``Risk-sensitive optimal control of quantum systems",
%\emph{Physical Review A}, Vol. 69, p. 032108, 2004.

\bibitem{James-et-al-2007}
James, M. R., Nurdin, H. I. \& Petersen, I. R. $H^\infty$ control of
linear quantum stochastic systems. \emph{IEEE Trans.
Automat. Control} \textbf{53}, 1787-1803 (2008).

\bibitem{Dong-and-Petersen-2009NJP}
Dong D. \& Petersen, I. R. Sliding mode control of quantum
systems. \emph{New J. Phys.} \textbf{11}, 105033 (2009).

%\bibitem{Dong and Petersen 2011Automatica}
%D. Dong and I.R. Petersen, ``Sliding mode control of two-level
%quantum systems", \emph{Automatica}, Vol. 48, pp.725-735, 2012.

%\bibitem{Qi 2013}
%B. Qi, ``A two-step strategy for stabilizing control of quantum systems with uncertainties,"
%{\em Automatica}, vol. 49, pp.834-839, 2013.

%\bibitem{Brif et al 2010}
%C. Brif, R. Chakrabarti and H. Rabitz, ``Control of quantum
%phenomena: past, present and future," {\em New Journal of Physics},
%Vol. 12, p.075008, 2010.

\bibitem{Wiseman-and-Milburn-2009}
Wiseman H. M. \& Milburn, G. J. {\em Quantum Measurement and
Control} (Cambridge University Press, Cambridge, England, 2010).

\bibitem{Gaitan-2008}
Gaitan, F. {\em Quantum Error Correction and Fault Tolerant Quantum Computing} (CRC Press, Boca Raton, Florida, USA, 2008).

\bibitem{Viola-et-al-1999}
Viola, L., Knill, E. \& Lloyd, S. Dynamical decoupling of open quantum systems. {\em Phys. Rev. Lett.} \textbf{82}, 2417-2421 (1999).

\bibitem{Khodjasteh-et-al-2010}
Khodjasteh, K., Lidar, D. A. \& Viola, L. Arbitrarily accurate dynamical control in open quantum systems. {\em Phys. Rev. Lett.} \textbf{104}, 090501 (2010).

\bibitem{Rabitz-et-al-2004}
Rabitz, H., Hsieh, M. M. \& Rosenthat, C. M. Quantum optimally controlled transition landscapes. {\em Science} \textbf{303}, 1998-2001 (2004).

\bibitem{Wilhelm-et-al-2007}
Sp\"{o}rl, A. K., \emph{et al.} Optimal control of coupled Josephson qubits.
{\em Phys. Rev. A} \textbf{75}, 012302 (2007).


\bibitem{Ginossar-et-al-2010}
Ginossar, E., Bishop, Lev S., Schuster, D. I. \& Girvin, S. M. Protocol for high-fidelity readout in the photon-blockade regime of circuit QED.
{\em Phys. Rev. A} \textbf{82}, 022335 (2010).


\bibitem{Wilhelm-et-al-2011}
Motzoi, F., Gambetta, J. M., Merkel, S. T. \& Wilhelm, F. K. Optimal control methods for rapidly time-varying Hamiltonians.
{\em Phys. Rev. A} \textbf{84}, 022307 (2011).

\bibitem{Wilhelm-et-al-2014}
Egger, D. J. \& Wilhelm, F. K. Adaptive hybrid optimal quantum control for imprecisely characterized systems.
{\em Phys. Rev. Lett.} \textbf{112}, 240503 (2014).

\bibitem{Chen-et-al-2013arXiv}
Chen, C.,  Dong, D., Long, R., Petersen, I. R. \& Rabitz, H. Sampling-based learning control of inhomogeneous quantum ensembles.
{\em Phys. Rev. A} \textbf{89}, 023402 (2014).

\bibitem{Martinis-et-al-2011PRL}
Bialczak, R. C.,  \emph{et al.} Fast tunable coupler for superconducting qubits.
{\em Phys. Rev. Lett.} \textbf{106}, 060501 (2011).

\bibitem{Martinis-et-al-2010PRA}
Pinto, R. A., Korotkov, A. N., Geller, M. R., Shumeiko, V. S. \& Martinis, J. M. Analysis of a tunable coupler for superconducting phase qubits.
{\em Phys. Rev. A} \textbf{82}, 104522 (2010).

\bibitem{Martinis-et-al-2013PRL}
Barends, R.,  \emph{et al.} Coherent Josephson qubit suitable for scalable quantum integrated circuits.
{\em Phys. Rev. Lett.} \textbf{111}, 080502 (2013).

\bibitem{Martinis-et-al-2014-1}
Chen, Y.,  \emph{et al.} Qubit architecture with high coherence and fast tunable coupling.
arXiv:1402.7367, quant-ph (2014).

\bibitem{Martinis-et-al-2014-2}
Geller, M. R., \emph{et al.} Tunable coupler for superconducting Xmon qubits: Perturbative nonlinear model.
arXiv:1405.1915, quant-ph (2014).

\bibitem{Nielsen-and-Chuang-2000}
Nielsen, M. A. \& Chuang, I. L. {\em Quantum Computation and Quantum
Information} (Cambridge University Press, Cambridge, England, 2010).

\bibitem{You-et-al-2003}
You, J. Q., Tsai, J. S. \& Nori, F. Controllable manipulation and entanglement of macroscopic quantum states in coupled charge qubits. {\em Phys. Rev. B} \textbf{68}, 024510 (2003).




%\bibitem{Dong and Petersen 2010IET}
%D. Dong, I.R. Petersen, ``Quantum control theory and
%applications: A survey," {\em IET Control Theory \& Applications},
%Vol.4, 2651-2671, 2010.

%\bibitem{Altafini and Ticozzi 2012}
%C. Altafini and F. Ticozzi, ``Modeling and control of quantum
%systems: an introduction," {\em IEEE Transactions on Automatic
%Control}, Vol. 57, No. 8, pp. 1898-1917, 2012.

\bibitem{Roslund-and-Rabitz-2009}
Roslund, J. \& Rabitz, H. Gradient algorithm applied to
laboratory quantum control. \emph{Phys. Rev. A} \textbf{79},
053417 (2009).
\newline

\end{thebibliography}
\end{document}